\documentclass
[preprint,english,aps,pra,superscriptaddress,nofootinbib]{revtex4-2}%
\usepackage{amsfonts}
\usepackage[T1]{fontenc}
\usepackage{amsmath}
\usepackage{amssymb}
\usepackage{babel}
\usepackage{graphicx}
\usepackage{subcaption}
\usepackage{xcolor}
\usepackage{lastpage}

\begin{document}

\title{Tunneling dynamics of the relativistic Schr\"{o}dinger/Salpeter equation}
\author{F. Daem}

\affiliation{Laboratoire de Physique Th\'eorique et Mod\'elisation, CNRS Unit\'e 8089, CY
	Cergy Paris Universit\'e, 95302 Cergy-Pontoise cedex, France}
	
	\author{A. Matzkin}

	\affiliation{Laboratoire de Physique Th\'eorique et Mod\'elisation, CNRS Unit\'e 8089, CY
		Cergy Paris Universit\'e, 95302 Cergy-Pontoise cedex, France}

\begin{abstract}
We investigate potential scattering and tunneling dynamics of a particle
wavepacket evolving according to the relativistic Schr\"{o}dinger equation
(also known as the Salpeter equation). The tunneling properties of the
Salpeter equation differ from those of the standard relativistic wave
equations (such as the Klein-Gordon or Dirac equations). In particular, the
tunneling solutions must be found by working in momentum space, given that the
equation in configuration space contains a pseudo-differential operator. The
resulting integral equations are derived and solved numerically for
wavepackets scattering on model potential barriers. The solutions are
characterized by the absence of Klein tunneling and an effect of the potential
on the fraction of the transmitted wavepacket that propagates outside the
light cone, a feature that has in the past been well-studied only for free propagation.

\end{abstract}
\maketitle

\newpage

\section{Introduction}

The relativistic Schr\"{o}dinger equation arises by plugging the canonically
quantized energy-momentum relation from relativistic classical mechanics into
the Schr\"{o}dinger equation, yielding%
\begin{equation}
i\hbar\partial_{t}\psi(t,x)=\sqrt{\hat{p}^{2}c^{2}+m^{2}c^{4}}\psi
(t,x)+V(\hat{x})\psi(t,x). \label{rse}%
\end{equation}
Eq. (\ref{rse}), also widely known as the Salpeter \cite{lucha} equation, or
the Newton-Wigner-Foldy equation \cite{horwitz}, or more prosaically as the
square-root Klein-Gordon equation is difficult to tackle both at the
conceptual and practical levels. Eq. (\ref{rse}) is not manifestly covariant
\cite{salpeter-ref} and the square-root gives rise to a non-local
pseudo-differential operator \cite{pseudo1}\ that makes the Salpeter equation
difficult to solve in position space. But on the other hand the Salpeter
equation has some attractive features -- it is a genuine single-particle
equation, which is not the case of the standard relativistic first-quantized
equations (e.g. the Klein-Gordon or the Dirac equations) and it has been
employed as a phenomenological tool to investigate low-energy relativistic
phenomena for spinless or spin-averaged particles, in particular the bound
states of hadrons \cite{meson-app-2003,meson-app-2006}. Analytical approaches
to investigate the bound state spectrum of the relativistic Schr\"{o}dinger
equation for different types (Coulomb, Dirac $\delta$, or harmonic oscillator)
of potentials have been developed
\cite{appli-quark,OH,coul-2009,deltas-wiese-2014,deltas-JPA,multiple-deltas-2017}%
. Free-wavepacket propagation has been investigated semi-analytically or numerically in configuration space \cite{usher,wiese,eckstein,pavsic,annalen,us-dynamics} or in mock phase-space \cite{wig1,wig2}.

In this paper we will investigate potential scattering and tunneling for the
Salpeter equation, a topic that has to our knowledge not received any
attention up to now: a particle wavepacket propagates freely towards a
potential barrier, at which point part of the wavepacket is reflected, and
part transmitted by tunneling. Free wavepacket propagation is known to display unusual properties
\cite{usher,wiese,eckstein,pavsic,annalen,us-dynamics}.\ In particular a small
fraction of the wavepacket propagates outside the light-cone
\cite{pavsic,pavsic2,us-dynamics}, a consequence of having a relativistic evolution
driven by a positive energy Hamiltonian \cite{hegerfeldt,kos}. Indeed, as it
is well-known in quantum field theory, relativistic propagators restricted to
the positive energy sector spill outside the light cone \cite{greiner-qft}: it
is only by including the contribution of the anti-particle sector that a
causal propagator can be obtained.

In order to find the solutions of an initial wavepacket impinging on a
potential barrier, Eq. (\ref{rse}) must be solved numerically in momentum
space, given that contrary to the Schr\"{o}dinger, Klein-Gordon or Dirac
equations, the square root prevents the obtention of an exact or approximate
basis to expand the solution inside the barrier in position space. We will
derive the integral equation in momentum space that needs to be solved in
Sec.\ \ref{sec2}. Wavepacket tunneling dynamics will be examined in
Sec.\ \ref{sec3}, where we will give numerical results as well as partial
analytical results in the limiting case of narrow barriers.\ We will discuss
these results (Sec.\ \ref{sec-disc}), focusing on the properties of the
transmitted wavepackets and their outside-the-light-cone
propagation.\ Concluding remarks will be given in Sec. \ref{sec-conc}.

\section{Salpeter equation in momentum space\label{sec2}}

\subsection{Free case}

We study the relativistic Schr\"{o}dinger equation in one spatial dimension.
The free solutions are readily obtained by working in momentum space
\cite{salpeter-ref,brau} since
\begin{equation}
\sqrt{-\hbar^{2}c^{2}\partial_{x}^{2}+m^{2}c^{4}}\psi(t,x)=\frac{1}{\sqrt
{2\pi\hbar}}\int dpE(p)e^{ipx/\hbar}\psi(t,p) \label{free}%
\end{equation}
with the energy $E(p)=\sqrt{p^{2}c^{2}+m^{2}c^{4}}$. The plane-waves of
positive energy $\exp\left(  ipx/\hbar-iE(p)t/\hbar\right)  $ fulfill Eq.
(\ref{rse}) with $V=0$. The time-dependent solution is then directly obtained
as \cite{salpeter-ref,wiese}
\begin{equation}
\psi(t,x)=\frac{1}{\sqrt{2\pi\hbar}}\int dpe^{ipx/\hbar}e^{-iE_{p}t/\hbar}%
\psi(0,p). \label{TE}%
\end{equation}
Typically we will interested in choosing $\psi(0,x)$ to be localized over a
compact support, picking $x_{0}$ and $p_{0}$ to be the average position and
momentum at $t=0$. This determines $\psi(0,p),$ and Eq.\ (\ref{TE}) is
evaluated by numerical integration in order to obtain the wavepacket evolution
\cite{annalen,us-dynamics}.

The non-local character of the pseudo-differential operator can be seen by
writing $\psi(t,p)$ in Eq. (\ref{free}) in terms of its inverse Fourier
transform, giving an integral with a kernel that is significant over distances
of a Compton wavelength $\lambda=\hbar/mc$ \cite{salpeter-ref}. A peculiar
feature of the wavepacket is that a fraction of the wavefunction leaks outside
the light-cone.\ While it is a mathematical fact \cite{hegerfeldt,heger-old}
that as soon as $t>0$, a wavefunction initially ($t=0$) localized over a compact support will be
non-zero everywhere on the real line, during a brief time interval the
OLC (outside the light-cone) fraction is significant \cite{us-dynamics}.

\subsection{Integral equation for a potential barrier}

Let us now include a finite potential barrier; the Salpeter equation is then
given by Eq. (\ref{rse}). Pure tunneling takes place when the wavepacket energy
entirely lies below the potential barrier.\ Standard relativistic wave
equations accept plane waves of both positive and negative energies as a basis
or even as solutions when $V(x)$ is a rectangular potential $V_{R}%
(x)=V_{0}\left[  \theta(x+L/2)-\theta(x-L/2)\right]  $. This is not the case
for the relativistic Schr\"{o}dinger equation due to the nonlocal character of
the pseudo-differential operator (\ref{free}). Note that a plane-wave solution inside the potential would require $E-V_{0}=\sqrt{p^{2}c^{2}+m^{2}c^{4}}$ to be a negative real
number, which is impossible.

Instead, basis functions must be obtained in momentum space, and the solution
Fourier transformed in order to follow the wavepacket evolve in position
space. Eq.(\ref{rse}) is readily obtained in momentum space as%
\begin{equation}
i\hbar\partial_{t}\psi(t,p)=\sqrt{p^{2}c^{2}+m^{2}c^{4}}\psi(t,p)+\frac
{1}{\sqrt{2\pi\hbar}}\int dp^{\prime}V(p-p^{\prime})\psi(t,p^{\prime})
\end{equation}
where%
\begin{equation}
V(p-p^{\prime})=\frac{1}{\sqrt{2\pi\hbar}}\int dxV(x)e^{-ix(p^{\prime}-p)/\hbar}
\label{vft}%
\end{equation}
is the Fourier transform of the potential. Separating the variables leads to
solutions of the form $\xi_{n}(t,p)=\exp\left(  -i\epsilon_{n}t/\hbar\right)
\phi_{n}(p)$ where $\phi_{n}(p)$ obeys%
\begin{equation}
\epsilon_{n}\phi_{n}(p)=E(p)\phi_{n}(p)+\frac{1}{\sqrt{2\pi\hbar}}\int
dp^{\prime}V(p-p^{\prime})\phi_{n}(p^{\prime}). \label{teie}%
\end{equation}
Finding the complete set of solutions $\phi_{n}(p)$ is necessary in order to
use these functions as a basis over which to expand the wavepacket through%
\begin{equation}
\psi(t,p)=\int dn\phi_{n}(p)e^{-i\epsilon_{n}t/\hbar}\int dp^{\prime}\phi
_{n}^{\ast}(p^{\prime})\psi(0,p^{\prime}). \label{timep}%
\end{equation}

Solutions of the integral equation (\ref{teie}) can be typically obtained in
closed form if $V(p-p^{\prime})$ is separable in $p$ and $p^{\prime}$
\cite{mex,lviv}, which is obviously not the case for a potential barrier
$V(x)$; for example for a rectangular barrier $V_{R}(x),$ Eq. (\ref{vft})
leads to $V_{R}(p-p^{\prime})\propto\sin\left[  (p-p^{\prime})L/(2\hbar)\right]
/\left(  p-p^{\prime}\right)  .$ An exception arises for an asymptotically
narrow barrier, in which case the integral equation (\ref{teie}) becomes
identical to the one for a Dirac delta $\delta(x)$ potential (see Sec.
\ref{sec-narrow} below). We must therefore in general solve Eq. (\ref{teie})
numerically.\ We will employ a simple discretization scheme particularly suitable for integral equations \cite{poljanin_handbook_1998}, by which the
integral equation is recast as an eigensystem problem
\begin{equation}
\epsilon_{n}\phi_{n}(p_{i})=\sum_{j=1}^{N}K_{ij}\phi_{n}(p_{j})\Delta p
\label{discrete}%
\end{equation}
where the kernel $K$ is given by%
\begin{equation}
K_{i,j}=\frac{E_{p_{i}}}{\Delta p}\delta_{i,j}+\frac{1}{\sqrt{2\pi\hbar}%
}V(p_{i}-p_{j}).
\end{equation}
$\Delta p=2\pi/(x_{max}-x_{min})$ is the step in the discretized momentum
space and $N$ the number of discretization points (we set $\hbar=1$ from now
on).\ Hence for a finite position interval $[x_{min},x_{max}]$, we have a step
$\Delta x$ with $x_{1}=x_{min}$, $x_{j}=x_{min}+j\Delta x$ and $x_{N}=x_{max}$.

This type of discretization scheme has already been applied to the Schrödinger equation in the past \cite{karr_numerical_2010}. We verified our numerical implementation against analytical solutions for the non-relativistic Schrödinger equation with a rectangular potential problem and for the free Salpeter equation, obtaining an excellent agreement.

\section{Tunneling dynamics\label{sec3}}

\subsection{General remarks}

We will choose a barrier of width $L$ and height $V_{0}$ centered at $x=0$,
with a wavepacket prepared at $t=0$ to the left of the barrier. The simplest
case is the rectangular potential%
\begin{equation}
V_{R}(x)=V_{0}\left[  \theta(x+L/2)-\theta(x-L/2)\right]  \label{rectb}%
\end{equation}
although for numerical solutions it is convenient to employ the smooth barrier%
\begin{equation}
V_{s}(x,\alpha)=\frac{V_{0}}{2}\left[  \tanh(\alpha(x+L/2))-\tanh\left(
\alpha(x-L/2)\right)  \right]  \label{smooth}%
\end{equation}
where $\alpha$ is the smoothness parameter.

The phenomenon of Klein tunneling, that is undamped oscillating waves inside
the barrier, is a well-known feature of the relativistic wave equations (see
e.g. \cite{PRA}). This relies on the fact that plane waves of momentum
$\sqrt{\left(  E-V_{0}\right)  ^{2}-m^{2}c^{4}}/c$ with $E>mc^{2}$ are
solutions of the Klein-Gordon or Dirac equations inside a rectangular barrier,
in particular for $V_{0}>2mc^{2}$, at which point the momentum is real and the
total energy $E-V_{0}$ is negative. Klein tunneling cannot happen for the
Salpeter equation, as we are restricted to positive energies. Only exponential-type
tunneling can be admitted inside the barrier, as is the case for the
non-relativistic Schr\"{o}dinger equation. Another difference with the usual
relativistic wave equations is that in the latter case, wave-packet
propagation is causal, including in the presence of potential barriers
\cite{math-new,us-new}; an initial wavepacket defined over a compact support
will therefore never leak outside the light cone originating from the
boundaries of the support. The Salpeter equation leaks outside the light-cone
for free-space propagation \cite{hegerfeldt,heger-old,pavsic,us-dynamics};
given that the non-local pseudo-differential operator in Eq. (\ref{free})
characterizes the equation irrespective of whether a potential is present, we
can expect that the transmitted wavepacket will also contain a fraction of the
amplitude propagating outside the light-cone.

\begin{figure}[ptb]
\centering
\includegraphics[width=0.4\linewidth]{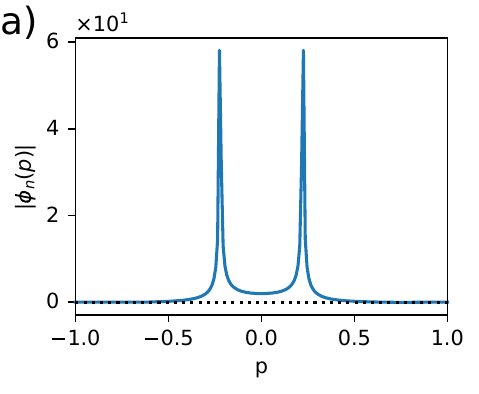}
\includegraphics[width=0.4\linewidth]{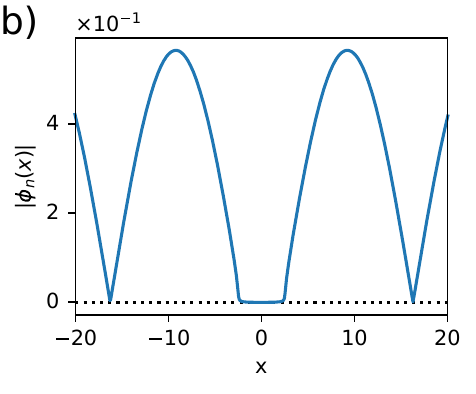}
\caption{Example
of an eigenfunction $\phi_{n}(p)$ obtained numerically by solving the integral
equation \eqref{discrete}. (a) Absolute value of $\phi_{n}(p)$. (b) Absolute value of the discrete Fourier transform of $\phi_{n}(p)$; in $x$-space the plateau
centered at $x=0$ corresponds to the region where the potential is
non-vanishing. The barrier parameters are $L=5$ and $V_{0}=20$ and $\phi
_{n}(p)$ corresponds to the eigenvalue $\epsilon_{n}\approx1.02$ (in natural
units: $\hbar=1,c=1,\lambda=\hbar/mc=1$).}%
\label{fig:eigen-typical}%
\end{figure}

The first step in setting up numerical calculations, independent of the
specific choice of the initial wavepacket, involves solving for the
eigenfunctions $\phi_{n}(p)$ of Eq. (\ref{teie}). A representative example of
an eigenfunction, as well as its Fourier transform in position space, is shown
in Fig. \ref{fig:eigen-typical} for a smooth potential $V_s(x,\alpha)$. This smooth potential tends to a rectangular one in the limit of large $\alpha$, and the value of $\alpha$ is chosen large enough so that the differences with a rectangular potential $V_R(x)$ are not noticeable. The eigenfunction that we obtain are the analog of the plane-waves
(in position space) that are the asymptotic building blocks (or exact building
blocks for a rectangular potential $V_{R}(x)$) of the standard wave equations
(although in a contrived way \cite{SR} for integer spin particles in case of superradiance).

\subsection{Numerical results}

We now show typical\ numerical results for an initial wavepacket of the form%
\begin{equation}
\label{eq:initwf}
\psi_{f}(0,x)=(\theta(x-x_{0}+\Delta_{x}/2)-\theta(x-x_{0}-\Delta_{x}%
/2))\cos^{8}\left[  \frac{\pi}{\Delta_{x}}\left(  x-x_{0}\right)  \right]
e^{ip_{0}x},
\end{equation}
corresponding to a wavefunction with mean position $x_{0}$ and momentum
$p_{0}$ defined over the compact support $[x_{0}-\Delta_{x}/2,x_{0}+\Delta
_{x}/2]$. The form $\cos^{8}$ is chosen to reduce the range in momentum space
over which the Fourier transform takes significant values and hence simplify
the resources needed for the numerical computations. Having solved
\footnote{We computed the solutions of the eigensystem in Python using the
\texttt{eigh} method in the \texttt{linalg} package of the \texttt{numpy}
library.} Eq. (\ref{discrete}) with the potential $V_{s}(x,\alpha)$, we use a
fast Fourier transform on Eq. (\ref{timep}) to obtain the time-dependent
wavepacket $\psi(t,x).$

\begin{figure}[ptb]
\centering
\includegraphics[width=0.32\linewidth]{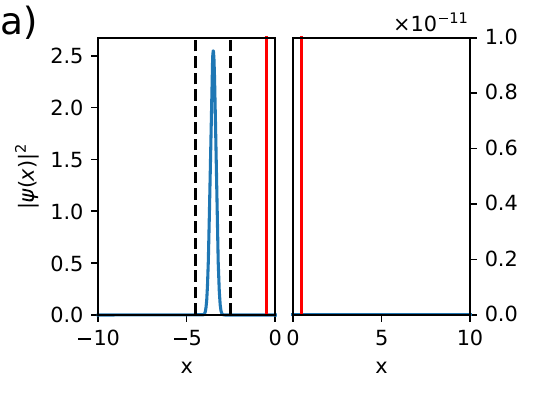}
\includegraphics[width=0.32\linewidth]{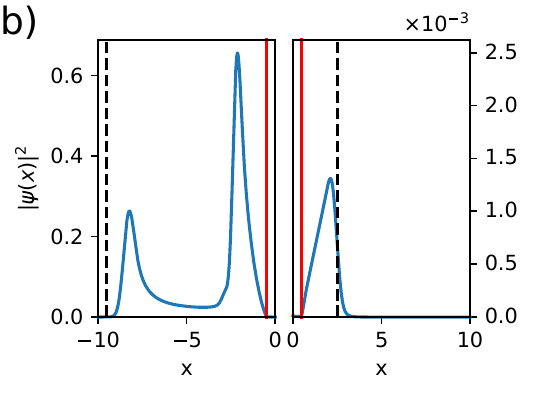}
\includegraphics[width=0.32\linewidth]{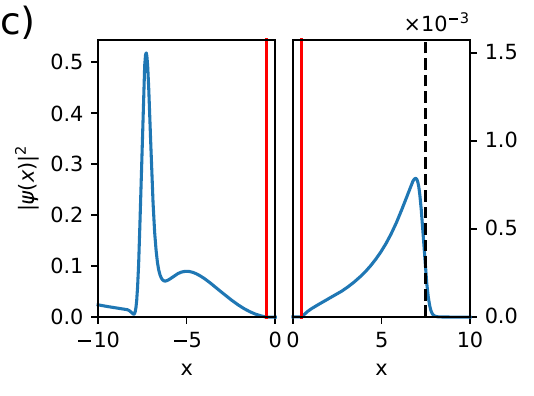}
\caption{Probability
density evolution of a wavepacket at $t=0$ (a), $t=5$ (b) and $t=10$ (c) in
natural units (as defined in Fig.\ \ref{fig:eigen-typical}). The initial
wavepacket is given by Eq. \eqref{eq:initwf} with $x_{0}=-3.5,p_{0}=1$ and
$\Delta_{x}=2$.\ The potential barrier depicted in red is given by Eq.
\eqref{smooth} with $V_{0}=20,$ $L=1$ and $\alpha=20$. The dashed lines in
(a) indicate the bounds of the compact support of the initial wavepacket; in
(b) and (c), the dashed lines indicate the position of the light cone
emanating from these bounds.}%
\label{fig:dynamic-full}%
\end{figure}

The tunneling dynamics of a typical wavepacket for a small value of $\Delta_{x}$ (of the
order of the Compton wavelength) is shown in Fig. \ref{fig:dynamic-full}. Part
of the initial wavepacket [Fig. \ref{fig:dynamic-full}(a)] moves toward the
left (corresponding to negative momenta components) while the other part
scatters on the barrier, giving rise to a reflected and a small transmitted
component (Fig. \ref{fig:dynamic-full}(b); beware of the different scale
employed on the panels featuring the transmitted wavepacket). A small fraction
of the transmitted component can be seen to move outside the light cone in
Fig. \ref{fig:dynamic-full}(b) and (c).

\begin{figure}[t]
\centering
\includegraphics[width=0.4\linewidth]{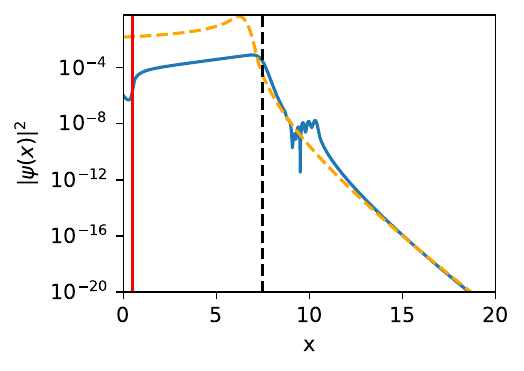}
\caption{Probability density in logarithmic scale of the transmitted wave packet
(blue line) at time $t=10$ (natural units) compared to the freely evolved
wavepacket (yellow dashed line). In both cases, a fraction of the probability
density lies beyond the light cone (dashed vertical line). The parameters of the initial wavepackets and the potential are the same as defined in Fig.\ \ref{fig:dynamic-full}.}%
\label{fig:transmitted}%
\end{figure}

Fig. \ref{fig:transmitted} focuses on the structure of the transmitted
wavepacket right after exiting the barrier, as compared to the same initial
wavepacket evolved freely. It can be seen that in both cases, part of the
amplitude lies outside the light-cone, but the tunneled wavepacket displays a
structure that is absent in the free case. This pattern varies with $L$ and $V_0$. We
will further analyze outside-the-light-cone propagation and compare it to the
case of free propagation below (Sec.\ \ref{sec-disc}), as well as the
properties of the transmitted peak as a function of the barrier and wavepacket parameters.

{}

\subsection{The narrow potential case ($L\rightarrow0$)\label{sec-narrow}}

Let us now take the rectangular potential $V_{R}(x)$ of Eq. (\ref{rectb}) and
assume the width $L$ is small. To first order in $L,$ the eigenvalue equation
(\ref{teie}) becomes%
\begin{equation}
\epsilon_{n}\phi_{n}(p)=\sqrt{p^{2}c^{2}+m^{2}c^{4}}\phi_{n}(p)+\frac{V_{0}%
L}{2\pi}\int dp^{\prime}\phi_{n}(p^{\prime}). \label{L0-teie}%
\end{equation}

\begin{figure}[ptb]
	\centering
	\includegraphics[width=0.4\linewidth]{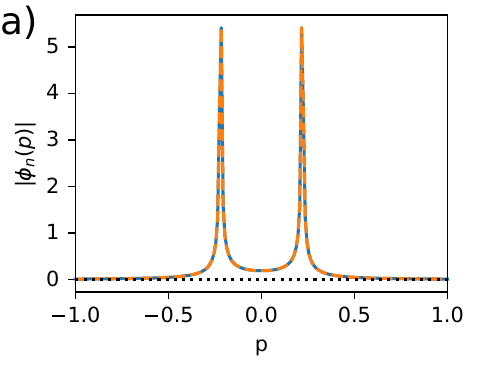}
	\includegraphics[width=0.4\linewidth]{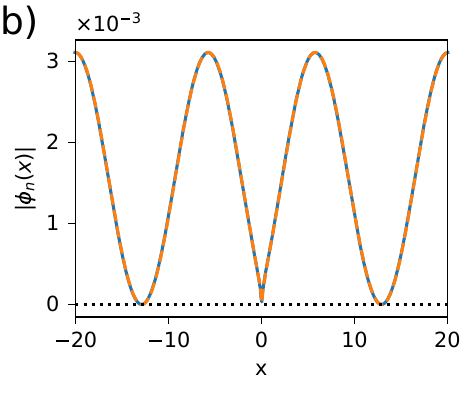}
	\caption{Modulus of the amplitude associated with an eigenfunction $\phi_{n}(p)$ with eigenvalue $\epsilon_{n}\approx1.02$ in
		natural units in $p$-space (a) and $x$-space (b) for a potential of very small width $L\approx 0.08$ (corresponding to the discretization step in $x$) and $V_0 L = 1$. The analytical result given by Eq. (\ref{L0-fi}) (yellow dashed lines) is computed on the same discretized points as the numerical solution (blue).}%
	\label{fig:L0-compare}%
\end{figure}

This is identical to the integral equation for the Salpeter equation in the
presence of a Dirac $\delta$ potential at $x=0$. Such equations have been
recently investigated \cite{deltas-wiese-2014,deltas-JPA,multiple-deltas-2017}%
, in particular in connection to bound states in an attractive Dirac
potential. Setting the integral $I_{n}=\int dp^{\prime}\phi_{n}(p^{\prime})$
to be an unknown constant, Eq. (\ref{L0-teie}) leads to%
\begin{equation}
\phi_{n}(p)=\frac{1}{\epsilon_{n}-\sqrt{p^{2}c^{2}+m^{2}c^{4}}}\frac
{V_{0}LI_{n}}{2\pi}. \label{L0-fi}%
\end{equation}
Note that attempting to integrate this equation in order to obtain $I_{n}$
leads to a divergent integral. It was suggested \cite{deltas-wiese-2014} to
renormalize the integral by employing dimensional regularization. Here the
knowledge of the energy dependence of $I_{n}$ would be needed in order to
integrate Eq. (\ref{timep}) and determine the time-dependent solution
$\psi(t,x)$ by Fourier transform. For a given value of $\epsilon_{n}$ however
we can rescale $\phi_{n}(p)$ with a global factor in order to compare Eq.
(\ref{L0-fi}) to our numerical solutions.\ An example is given in Fig.
\ref{fig:L0-compare}, showing a good agreement. While we have found it
impossible to obtain analytical solutions by integrating Eq. (\ref{timep}) and
Fourier transforming it, we point out that the solutions in the case of small
$L$ present are similar to the one for larger values of $L$ shown in Fig.
\ref{fig:eigen-typical}, save for the absence of the plateau centered at $x=0$
in $\phi_{n}(x)$ that precisely characterizes the width of the barrier.

\begin{figure}[ptb]
	\centering
	\includegraphics[width=0.4\linewidth]{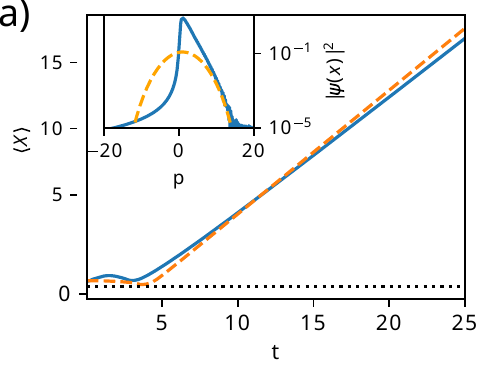}
	\includegraphics[width=0.4\linewidth]{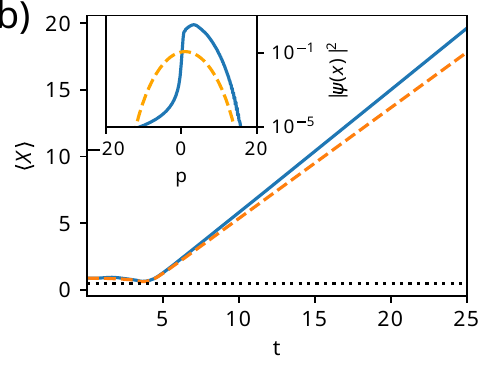}
	\caption{Mean
		value of the position of the transmitted wave packet for the tunneling case
		(blue line) and the free case (dashed yellow line). Each figure includes an associated inset where the momentum distribution of the wave packets is displayed on a logarithmic scale. The parameters and units employed in (a) are the same as in Fig.\ \ref{fig:dynamic-full}. In (b), the same parameters and units are used, except for $L=1$ and $V_0=1$.}%
	\label{fig:transmitted-average}%
\end{figure}

\section{Discussion\label{sec-disc}}

\begin{figure}[t]
	\centering
	\includegraphics[width=0.4\linewidth]{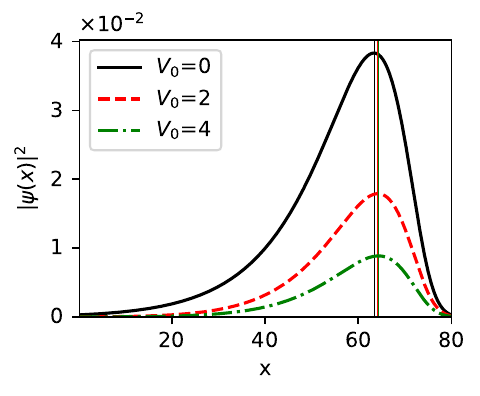}
	\caption{Transmitted wavepackets at time $t=100$ in
		natural units (as defined in Fig.\ \ref{fig:eigen-typical}) for $V_0=2$ (red dashed line), $V_0=3$ (green dashed line) and the free case $V_0=0$ (black line). The initial wave packet is defined in Eq. \eqref{eq:initwf} with $x_{0}=-22.25,p_{0}=1$ and $\Delta_{x}=20$. The colored vertical lines correspond to the $x$ value at which the corresponding packet is maximal.}
	\label{fig:transmitted-packet-average}
\end{figure}

We have seen that broadly speaking tunneling in the relativistic
Schr\"{o}dinger equation appears to follow the familiar pattern known from
non-relativistic tunneling \cite{stun1,stun2}. Most of the amplitude is reflected and only a wavepacket with a tiny amplitude is transmitted. One well-known feature of non-relativistic tunneling is
that the average of the position of the transmitted peak is advanced, during 
a transient time interval, relative
to the average position of the same wavepacket propagating freely.
Then the tunneled wavepacket accelerates or decelerates relative to the freely propagating one. The first feature is usually associated with wave packet reshaping.
The second one is due to the dependence of the transmission coefficient on energy, effectively filtering some momentum components of the wavefunction depending on the parameters of the potential barrier. This is
also the case here, as illustrated in Fig. \ref{fig:transmitted-average}. Fig. \ref{fig:transmitted-average} (a) shows an example corresponding to a wave packet transmitted earlier than in the free case but with a slower average momentum, while Fig. \ref{fig:transmitted-average} (b) corresponds to a transmitted wave packet with faster contributions.
The evolution of the mean position of the transmitted wavepacket right after exiting the barrier, as a function of the barrier width or height, also follows the same pattern known for Schr\"{o}dinger tunneling \cite{stun2}. An example displaying  the transmitted wavepacket as a function of the potential height is shown in Fig. \ref{fig:transmitted-packet-average}.

\begin{figure}[ptb]
\centering
\includegraphics[width=0.4\linewidth]{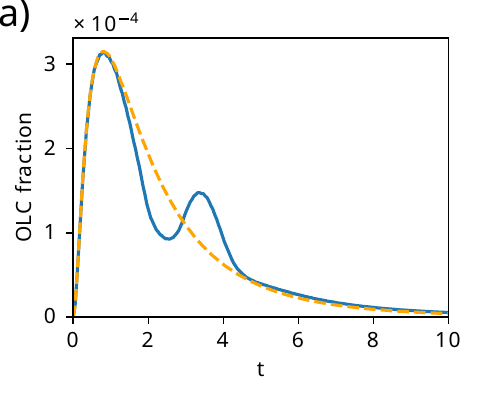}
\includegraphics[width=0.4\linewidth]{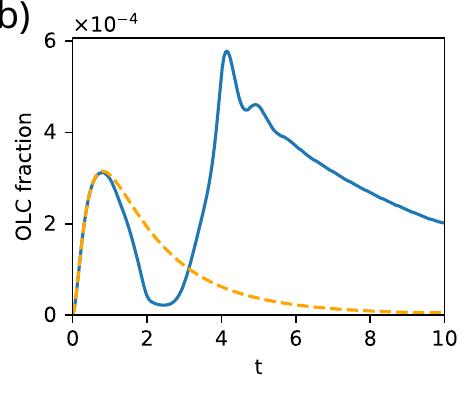}
\includegraphics[width=0.4\linewidth]{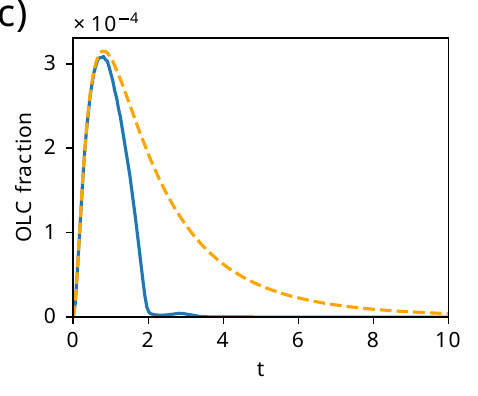}
\caption{Outside
the Light-Cone (OLC) fraction of the wavepacket (blue line)
compared to the free case (dashed yellow line) as a function of time. At short times the behavior
is similar, but in the tunneling case a second peak in the OLC fraction
appears, whose form and amplitude depend on $L$ and $V_0$. Three typical behaviors are presented with (a): $L=1$ and $V_0=1$, (b): $L=2$ and $V_0=5$, (c): $L=10$ and $V_0=20$, where the units and other parameters are as defined in Fig.\ \ref{fig:dynamic-full}.}%
\label{fig:olc-frac}%
\end{figure}

As mentioned above, a specific feature of Salpeter wavepackets is their propagation outside
the light-cone \cite{pavsic,hegerfeldt,kos}. The fraction of the wavepacket
propagating outside the light-cone (OLC) was quantified previously for free
wavepackets \cite{us-dynamics}. It is interesting to look at the effect of the
barrier on this type of propagation. An illustration is provided in  Fig. \ref{fig:olc-frac}, that shows the 
time-dependence of the OLC fraction of the right edge of the wavepacket in a typical situation. 
In the free case, OLC propagation appears as a transient effect that reaches a maximum at short times and then falls off gradually. In the tunneling case, the OLC fraction evolves at first like the one for the free wavepacket (essentially as the wavepacket is traveling toward the barrier) but an additional structure appears in the form of a second peak (or a plateau, depending on $L$). We hypothesize that this additional structure might result
from superluminal waves of the first peak reflecting
inside the barrier and then transmitted before being caught up by the light cone. This is supported by the fact that for larger barrier widths, the second structure disappears. Hence tunneling appears to have an effect on the OLC fraction only for small barrier widths.

\begin{figure}[ptb]
\centering
\includegraphics[width=0.4\linewidth]{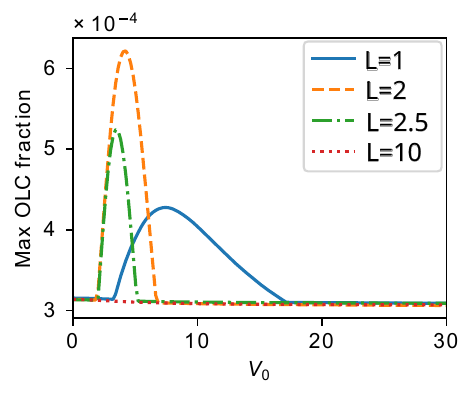}
\caption{Global maximum value of the
outside the light cone (OLC) fraction of the transmitted wavepacket as a
function of $V_{0}$ for different barrier widths: $L=1$ (blue line), $L=2$
(orange dashed line), $L=2.5$ (green dashed line) and $L=10$ (red dotted line), for the same initial
wavepacket having parameters $x_{0}=-(L/2+3)$, $p_{0}=1,$ $\Delta_{x}=2$. Numbers
are given in natural units $c=\hbar=1,$ $\lambda=\hbar/mc=1$.}%
\label{fig:olc-param}%
\end{figure}

We have also determined the behavior of the maximum value of the OLC fraction (irrespective of the time at which this maximum takes place) as
a function of $V_{0}$ for different widths $L$.\ Results for a given initial
wavepacket (identical in all cases considered) are displayed in Fig.
\ref{fig:olc-param}. This figure quantifies the relative importance of the second OLC peak shown in  Fig. \ref{fig:olc-frac}, due to
the presence of barrier, relative to the first peak, already present in free propagation (the free case OLC is visible in Fig. \ref{fig:olc-param}  by looking at the $V_{0}\rightarrow0$ limit).  The effect of the barrier on the OLC fraction of the transmitted wavepacket is only substantial for small values of $L$ (for $L=10$ natural units, the OLC fraction is seen to be identical to the one for free propagation) and for ``moderate'' values of $V_0$. This is consistent with the idea that the OLC structure due to tunneling is due to reflected waves (that are caught up by the light cone for larger values of $L$), although it does not explain the fact that for a given small value of $L$, there is an value of  $V_0$ maximizing the OLC fraction (understandably, for large values of $V_0$, the reflected waves are too small to contribute to the OLC fraction significantly).

\section{Conclusions\label{sec-conc}}

We have investigated the tunneling dynamics for a particle scattering on a
potential barrier in the context of the relativistic Schr\"{o}dinger equation.
While in the non-relativistic case or for standard relativistic equations
potential tunneling can be handled approximately or asymptotically with
plane-waves, the fact that the Salpeter equation contains a square-root
operator renders the equation difficult to handle analytically beyond the free
case \cite{usher,salpeter-ref}.\ We have therefore undertaken a numerical
approach by computing the solutions of the corresponding integral equation in
momentum space.

While the Salpeter equation does not describe the fundamental physics of spin-0 particles, it should be recalled that the correct relativistic equation (the Klein-Gordon equation) when written in the Foldy-Wouthuysen (FW) representation takes the form in free space of two uncoupled Salpeter equations. The barrier potential couples
the two Salpeter equations, so that we can expect our present results to be an adequate approximation of the Klein-Gordon tunneling dynamics (in the FW representation) in the limit of weak couplings (low barriers).

These results indicate that the tunneling dynamics is similar to
standard non-relativistic tunneling, or for that matter to the
tunneling process for relativistic equations when the potential is far below
the supercritical threshold (as expected, the Salpeter equation does not display Klein tunneling in that case). The peculiar outside-the-light-cone
propagation\ of the relativistic Schr\"{o}dinger wavepackets was seen to be
modified by the tunneling process only in the case of rather narrow barriers, the overall effect remaining however very small. It would be useful to develop analytical methods giving approximate transmission amplitudes as a function of the barrier parameters.

\end{document}